\documentclass[11pt, a4paper]{article}
\usepackage{float}
\usepackage{indentfirst} 
\usepackage[top=3cm, bottom=3cm, left=2.0cm, right=2.0cm]{geometry}

\usepackage[utf8]{inputenc}
\usepackage{graphicx}
\usepackage{subfig}
\usepackage{xcolor}
\usepackage{amsfonts}
\usepackage{amsmath}
\usepackage{amssymb}
\usepackage{authblk}
\usepackage{physics}
\usepackage{hyperref}
\hypersetup{
            colorlinks=true,
            linkcolor=blue,
            urlcolor=violet,
            citecolor=violet,
            }

\usepackage{cite}

\newcommand{\ti}[1]{\tilde{#1}}
\newcommand{\tp}{\tilde{p}}
\newcommand{\tir}{\tilde{r}}

\newcommand{\of}[1]{ \! \left( #1 \right) }

\begin{document}
\begin{titlepage}
\title{\LARGE Revisiting a family of five-dimensional charged, rotating black holes}
\author[1,3]{Marcos R. A. Arcod\'ia\thanks{marcodia@iafe.uba.ar}}
\author[2]{Gaston Giribet\thanks{gaston.giribet@nyu.edu}}
\author[3]{Juan Laurnagaray\thanks{jlaurna@gmail.com}}

\affil[1]{Instituto de Astronom\'{\i}a y F\'\i sica del Espacio (IAFE, CONICET-UBA), Casilla de Correo 67, Sucursal 28, 1428 Buenos Aires, Argentina.}
\affil[2]{Center for Cosmology and Particle Physics, Department of Physics, New York University, 726 Broadway, New York, 10003, USA.}
\affil[3]{Departamento de F\'\i sica, Facultad de Ciencias Exactas y Naturales, Universidad de Buenos Aires, Ciudad Universitaria, Pabell\'on I, 1428 Buenos Aires, Argentina.}

\date{}

\maketitle
\thispagestyle{empty}

\begin{abstract}
In the absence of a higher-dimensional analogue to the Kerr-Newman black hole, 5-dimensional Einstein-Maxwell theory with a Chern-Simons term has become a natural setting for studying charged, stationary solutions. A prominent example is the Chong-Cveti\v{c}-L\"{u}-Pope (CCLP) solution, which describes a non-extremal black hole with electric charge and two independent angular momenta. This solution has been widely studied, and generalizations have been proposed. In this paper, we revisit a large family of five-dimensional black hole solutions to Einstein-Maxwell-Chern-Simons (EMCS) field equations, which admits to be written in terms of a generalized Pleba\'{n}ski-Demia\'{n}ski ansatz and includes the CCLP and the Kerr-NUT-Anti-de Sitter solutions as particular cases. We show that the complete family can be brought to the CCLP form by means of a suitable coordinate transformation and a complex redefinition of parameters. Then, we compute the conserved charges associated to the CCLP form of the metric by analyzing the near-horizon asymptotic symmetries. We show that the zero-mode of the near-horizon charges exactly match the result of the Komar integrals.
\end{abstract}
\end{titlepage}

\newpage

\section{Introduction}

Einstein-Maxwell-Chern-Simons (EMCS) theory in five spacetime dimensions provides an excellent model for investigating stationary, charged, and back-reacting solutions, including rotating black holes. The absence of a five-dimensional analogue of the Kerr-Newman solution in general relativity, together with the variety of geometries in dimensions higher than four, hampers the analytical study of charged, rotating solutions in more than four dimensions. A special case where this is possible is precisely Einstein gravity coupled to an Abelian gauge field with the addition of a Chern-Simons term. The theory thus defined has been considered in the literature to study various aspects of non-extremal black holes in higher dimensions. For example, the derivation of logarithmic corrections to the Bekenstein-Hawking entropy formula from the Kerr/CFT correspondence has been considered as a practical example \cite{Pathak:2016vfc}.

Although complicated enough, the field equations of the  EMCS theory in five dimensions have been explored and a number of explicit solutions have been found. The catalog of solutions known so far is interesting, including multiple angular momenta, electric charges, a NUT-like parameter, and extra parameters that lead to different asymptotic behaviors. The purpose of this work is to study the main families of known solutions, the relations among them, their global features, their conserved charges and thermodynamic variables.

A large family of metrics that solve the EMCS field equations in five dimensions was obtained in \cite{Ferraro}. This family describes spacetimes with electric charge, mass, a NUT-like parameter, independent rotations in two orthogonal planes, and three additional parameters. It includes, as particular cases, the well-known Chong-Cveti\v{c}-L\"{u}-Pope (CCLP) solution \cite{cclp} as well as the Kerr-NUT-(A)dS spacetime \cite{knads}. In \cite{Arcodia:2021fge}, it was shown that the solutions derived in \cite{Ferraro} are the most general electrovacuum spacetimes admitting a double-extended Kerr-Schild form. Therefore, studying the physical interpretation of the parameters of such solutions is important. 

A key challenge in analyzing the physical properties of such a multi-parameter solution is identifying the appropriate orthonormal frame that would permit to compute its conserved charges. The presence of a NUT-like charge, along with the presence of new parameters, hampers the detection of global pathologies that could preclude the correct calculation of the charges. One possible strategy to circumvent this technical obstruction is to look for the a coordinate transformation that renders the metric manifestly asymptotically locally flat, de Sitter (dS), or Anti-de Sitter (AdS). Another strategy would be to investigate whether a coordinate redefinition suffices to eliminate the NUT-like parameter --it has been demonstrated for the case of the Kerr-NUT-(A)dS metric that in five dimension such parameter can in principle be removed \cite{kadsnut}--. Here, we will explore these strategies for the family of metrics found in \cite{Ferraro}. This will enable us to show that such solutions turn out to be equivalent to the CCLP metrics via a complex redefinition of the parameters and a suitable local change of coordinate. This observation is of great help in investigating the physical properties and geometric interpretation of the results of \cite{Ferraro}. For example, by relating the solutions in \cite{Ferraro} to those in \cite{cclp} one can identify the right asymptotics and, consequently, calculate the conserved charges. Since the CCLP solution has been extensively discussed in the literature  \cite{Cvetic:2010jb,
 Kinney:2005ej,
Cabo-Bizet:2018ehj,
Dolan:2013ft,
Maldacena:2008wh,
Kunduri:2007vf,
Cvetic:2010mn,
Chow:2008dp,
Kunduri:2006ek,
Gutowski:2004ez} --especially recently\cite{Gray:2024qys, Deshpande:2024itz, Zhao:2024ehs, Larsen:2024fmp, Cabo-Bizet:2024gny, Ma:2024ynp, Iliesiu:2024cnh, Cano:2024tcr, Hassaine:2024mfs, Deshpande:2024vbn, Colombo:2025ihp, David:2025pby, Barrientos:2025abs, Hale:2024zvu, Dias:2024edd, Horowitz:2024kcx, Ezroura:2024xba, Budzik:2023vtr, Mao:2023qxq, David:2023gee}--, establishing its connection with other families of solutions is interesting \cite{Ferraro}.

The computation of the conserved charges associated to these families of solutions to EMCS field equations can also be performed in another way, without dealing with its far asymptotic behavior. This is achieved by resorting to the near-horizon method developed in \cite{Donnay:2015abr, Donnay:2016ejv}. This method has been applied to different black holes, mostly in 4-dimensional Einstein theory \cite{cosmological, cmetric}, Einstein-Maxwell theory \cite{memoryeffect, meissner, sashalucho} and Einstein-Yang-Mills \cite{kacmoody}. The method can also be applied in dimension 5 \cite{Mei:nearhorizon}, including the case of horizons with non-trivial topology \cite{ringo}. Here, we will apply it in the case a CS term is also present. We will perform the near-horizon computation of the charges and compare the result with the Komar integral, finding exact agreement.

The paper is organized as follows: In section \ref{5Delectrovacuum}, we review the family of solutions found in \cite{Ferraro} as well as some particular cases. In section \ref{sec:equivalence_cclp}, we propose a coordinate change that results to be useful to show the relation between the solutions in \cite{Ferraro} and the CCLP metrics \cite{cclp}. We also discuss the existence of conical singularities and the possibility of closed timelike curves. The correct identification of the global properties of the asymptotic geometry turns out to be important to compute the Komar integrals and obtain the conserved charges. In section \ref{sec:near_horizon}, we perform an alternative computation of the conserved charges resorting to the near-horizon formalism applied to the EMCS theory. We compute the charges associated to the CCLP metrics in this way, obtaining results in agreement with the results obtained by different methods. In section \ref{sec:conclusions}, we summarize the results.

\section{Five-dimensional electrovacuum geometries}\label{5Delectrovacuum}

In this section, we will introduce the family of geometries we will be concerned with. Consider the five dimensional  EMCS action given by
\begin{equation}
    S = \int \left[ k_\text{E} (R + 12 \lambda) \star 1 +  k_\text{M} \, F \wedge \star F + k_\text{CS} \, F \wedge F \wedge A \right] \, ,
    \label{eq:emcs_action}
\end{equation}
where $k_\text{E}$, $k_{\text{M}}$ and $k_{\text{CS}}$ are the Einstein, Maxwell and Chern-Simons coupling constants respectevly. $\lambda$ is related to the cosmological constant\footnote{Because of the minus sign in the cosmological constant relation AdS correspond to $\lambda > 0$.} by $\Lambda = - 6 \lambda$. The equations of motion can be obtained by varying \eqref{eq:emcs_action} with respect to the fields; namely
\begin{equation}
    R_{\mu \nu} - \frac{1}{2} g_{\mu \nu}(R + 12 \lambda) + \frac{k_{\text{M}}}{k_{\text{E}}} {T}_{\mu \nu} = 0 \, , \qquad  d \star F + \frac{3}{2} \frac{k_\text{CS}}{k_\text{M}} F \wedge F = 0 \, ,
    \label{eq:emcs_eom}
\end{equation}
with the stress-energy tensor
\begin{equation*}
    {T}_{\mu \nu} = {F_{\mu}}^{\lambda} F_{\nu \lambda} - \frac{1}{4} g_{\mu \nu} F_{\alpha \beta} F^{\alpha \beta} \, .
\end{equation*}

We will study the family of solutions of \eqref{eq:emcs_eom} found in reference \cite{Ferraro}, that can be casted in an extended Pleba\'{n}ski-Demia\'{n}ski \textit{Ansatz} with coordinates $(t,\phi,\psi,r,p)$; namely
\begin{equation}
    \mathbf{g} = -\frac{Y\of{r}}{p^2 + r^2} \omega^0 \otimes \omega^0 + \frac{X\of{p}}{p^2 + r^2} \omega^1 \otimes \omega^1 + \frac{1}{p^2 r^2} \Omega^2 \otimes \Omega^2 + (p^2 + r^2) \left( \frac{dr \otimes dr}{Y\of{r}} + \frac{dp \otimes dp}{X\of{p}} \right) \, ,
    \label{eq:metric_rafael}
\end{equation}
with spacelike coordinates $r, p$, a timelike coordinate $t$ and two angular coordinates $\phi , \psi$, and with the differential forms
\begin{equation}
    \begin{aligned}
        \omega^0 & \equiv \frac{(1 - p^2 \lambda) \, dt}{\Xi_a \, \Xi_b} - \frac{a (a^2 - p^2) \, d\phi}{(a^2 - b^2) \, \Xi_a} - \frac{b (b^2 - p^2) \, d\psi}{(b^2 - a^2)\, \Xi_b} \, , \\ \\
        \omega^1 & \equiv \frac{(1 + r^2 \lambda) \, dt}{\Xi_a \, \Xi_b} - \frac{a (a^2 + r^2) \, d\phi}{(a^2 - b^2) \, \Xi_a} - \frac{b (b^2 + r^2) \, d\psi}{(b^2 - a^2) \, \Xi_b} \, , \\ \\
        \Omega^2 & \equiv -\frac{a b (1 + r^2 \lambda)(1 - p^2 \lambda) \, dt}{\Xi_a \, \Xi_b} + \frac{b(a^2 + r^2)(a^2 - p^2) \, d\phi}{(a^2 - b^2) \, \Xi_a} \\
        & \hspace{115pt} + \frac{a(b^2 + r^2)(b^2 - p^2) \, d\psi}{(b^2 - a^2)\, \Xi_b} - \frac{p^2 r^2}{(p^2 + r^2)} \left ( \mathcal{Y}\of{r} \omega^0 + \mathcal X\of{p} \omega^1 \right )
    \end{aligned}
    \label{eq:rafael_base}
\end{equation}
and\footnote{Notice a sign change in $\lambda$ with respect to the definitions of $X\of{p}$ and $Y\of{r}$ in (9) and (10) of \cite{Ferraro}.}
\begin{equation}
    \begin{aligned}
        \mathcal X\of{p} & = \mu_x \frac{Q}{p^2} \, , \qquad \mathcal Y\of{r} = \mu_y \frac{Q}{r^2} \, , \qquad  \Xi_a = 1 - a^2 \lambda, \qquad \Xi_b = 1 - b^2 \lambda \, , \\
        X\of{p} & = - \frac{(a^2 - p^2)(b^2 - p^2)(1 - \lambda{p^2})}{p^2} + \alpha p^2 + 2 n - \mu_x Q \, \frac{\mu_x Q + 2 a b}{p^2}\\
        & =\lambda p^4 -  \left(1-\alpha+\lambda(a^2+b^2)\right)p^2+(a^2+b^2+\lambda a^2 b^2+2n)-\frac{\left(a b + Q \mu_{x}\right)^2}{p^2} \\ 
        Y\of{r} & = \frac{(a^2 + r^2) (b^2 + r^2) (1 + \lambda{r^2})}{r^2} - \alpha r^2 - 2 m + \mu_y Q \,\frac{\mu_y Q + 2 a b}{r^2}\\
        & =\lambda r^4 + \left(1-\alpha+\lambda(a^2+b^2)\right)r^2+(a^2+b^2+\lambda a^2 b^2-2m)+\frac{\left(a b + Q \mu_{y}\right)^2}{r^2}
    \end{aligned}
    \label{eq:rafael_metric_functions}
\end{equation}
The electromagnetic potential reads
\begin{equation}
    A = \frac{\chi \, Q}{p^2 + r^2} \omega^0 \, .
    \label{eq:ferraro_em_potential}
\end{equation}

The metric has three constants\footnote{The reason why we make a distinction between these constats and parameters will be discussed along the paper.}, $\alpha$, $\mu_x$  and $\mu_y$, and five parameters, the rotation parameters $a, b$ (without loss of generality we will take $a < b$), the parameter $\lambda$ (which is related to the cosmological constant by $\Lambda=-6\lambda$), the mass parameter $m$, the NUT-like charge $n$, and the electric charge parameter $Q$. For the field equations to hold, the parameter $\chi$ must satisfy the equation
\begin{equation}
    \chi=\pm\frac{2k_M}{3k_{CS}}(\mu_y-\mu_x) \, .
\end{equation}
Here we will take $\chi=\pm\frac{2k_M}{3k_{CS}}$, so the condition $\mu_y-\mu_x=\pm{1}$ must hold.

The solution presented above includes the CCLP metric \cite{cclp} as the particular case $\mu_x = 0$, $\mu_y = 1$, $n = 0$, $\alpha=0$, $\lambda = g^2$, $\chi = \sqrt{3}$, with the coordinate definition $p^2 = a^2 \cos^2\theta + b^2 \sin^2\theta$. We will generically consider the so-called under-rotating case $\Xi_a, \Xi_b > 0$, and asymptotically AdS spacetimes, which correspond to $\lambda > 0$.

The metric has Lorentzian signature provided $X(p)>0$, so the rank of the variable $p$ will be determined by that condition. For $\alpha=0$, $n=0$, $\mu_{x}=0$, $X(p)$ takes a factorized form and consequently $p\in(a,b)$ for any sign of $\lambda$ as long as we are in the under-rotating case. We will analize the rank of $p$ for $n\neq{0}$, $\alpha\neq{0}$ and $\mu_x\neq{0}$ in the following sections.

The ranges of $\phi$ and $\psi$ are also easily determined for the cases where $\alpha=n=\mu_{x}=0$. This is because in the coordinate system we are using, it is clearly seen that the metric is asymptotically (A)dS or Minkowski (if $\lambda=0$) in oblate spheroidal coordinates. If the parameters $n$, $\alpha$ or $\mu_x$ do not vanish, the case is more complicated and will be the subject of discussion in this article. The determination of the correct domain of the $\phi$-$\psi$ pair of variables requires to factor the function $X(p)$ and analize the asymptotic behavior of the metric and conical singularities.

One of the questions we want to answer is that of the physical meaning of all the parameters of the solution reviewed above, as well as its connection with other solutions of the Einstein-Maxwell-Chern-Simos theory reported in the literature. This will require, on the one hand, to study the global properties of the solution. On the other hand, the calculation of the conserved charges associated to the geometries is important. As for the relation of the solution reviewed above and other known solutions of the theory, we have already discussed the connection to the CCLP solution, which appears as a particular case. The question remains as to whether the more general case, with more parameters turned on, can also be written in the CCLP form and, if so, what the precise relation is. We discuss this in detail in the next section. 

\section{Equivalence with the CCLP metric}\label{sec:equivalence_cclp}

If in the solution above we perform the coordinate change
\begin{eqnarray}\notag
        &&t \to t +\left(a^2+b^2\right)\psi+a b (a b +Q \mu_{x}) \, \phi \, , \qquad
        \phi \to (a b+ Q \mu_{x}) b \phi + a \lambda t +\left(b^2 \lambda +1\right)a \psi \, , \\ \label{eq:newcoords} \\
        && \psi \to \left(a^2 \lambda +1\right) b \psi+b \lambda  t +(a b+Q\mu_{x}) a \phi \notag \, ,
\end{eqnarray}

\noindent the differential forms $\omega^0, \ \omega^1$ and $\Omega^2$ take the form
\begin{equation}
    \begin{aligned}
    &\omega^{0}=dt - p^2 d\psi \ , \quad \omega^{1}=dt+ r^2 d\psi \, , & \\
    &\Omega^{2}=-\left[(a b +Q \mu_{x})+\frac{Q p^2}{p^2+r^2}\right]dt+p^2 r^2 (a b + Q \mu_{x})d\phi-\left[\frac{Qp^4}{p^2+r^2}+(p^2-r^2)(a b + Q \mu_{x})\right]d\psi \, .
    \end{aligned}
\end{equation}
Now, if we take the coordinate transformation
\begin{eqnarray} \label{eq:newcoordsbis}
    r=\kappa \tilde{r} \ , \ p=\kappa{\tilde{p}} \ , \ t= \frac{\tilde{t}}{\kappa} \ , \ \phi=\frac{\tilde{\phi}}{\kappa^5} \ , \ \psi=\frac{\tilde{\psi}}{\kappa^3}    
\end{eqnarray}
and simultaneously rescale
\begin{equation}
    \tilde{X}(\tilde{p})=\frac{X(\kappa{p})}{\kappa^4} \ , \     \tilde{Y}(\tilde{r})=\frac{Y(\kappa{r})}{\kappa^4} \ , \ \tilde{Q}=\frac{Q}{\kappa^3} \ ,
\end{equation}
and define parameters $\tilde{a}, \ \tilde{b}$ as
\begin{equation}
    \tilde{a}\tilde{b}=\frac{a b + Q\mu_{x}}{\kappa^3}
\end{equation}
the metric takes the following form
\begin{equation}
    \mathbf{g} = -\frac{\tilde{Y}\of{\tilde{r}}}{\tilde{p}^2 + \tilde{r}^2} \tilde{\omega}^0 \otimes \tilde{\omega}^0 + \frac{\tilde{X}\of{\tilde{p}}}{\tilde{p}^2 + \tilde{r}^2} \tilde{\omega}^1 \otimes \tilde{\omega}^1 + \frac{1}{\tilde{p}^2 \tilde{r}^2} \tilde{\Omega}^2 \otimes \tilde{\Omega}^2 + (\tilde{p}^2 + \tilde{r}^2) \left( \frac{d\tilde{r} \otimes d\tilde{r}}{\tilde{Y}\of{\tilde{r}}} + \frac{d\tilde{p} \otimes d\tilde{p}}{\tilde{X}\of{\tilde{p}}} \right) \, ,
    \label{eq:metric_tildedcoords}
\end{equation}
with
\begin{eqnarray}
    &\tilde{\omega}^{0}=d\tilde{t} - \tilde{p}^2 d\tilde{\psi} \ , \quad \tilde{\omega}^{1}=d\tilde{t}+ \tilde{r}^2 d\tilde{\psi} \  ,\\ \notag \\
    &\tilde{\Omega}^{2}=-\left(\tilde{a}\tilde{b}+\frac{\tilde{Q}\tilde{p}^2}{\tilde{p}^2+\tilde{r}^2}\right)d\tilde{t}+\tilde{p}^2 \tilde{r}^2 \tilde{a}\tilde{b} d\tilde\phi-\left(\frac{\tilde{Q}\tilde{p}^4}{\tilde{p}^2+\tilde{r}^2}+(\tilde{p}^2-\tilde{r}^2)\tilde{a}\tilde{b}\right)d\tilde{\psi}\ ,
\end{eqnarray}
which is the exact form of the ansatz for $\mu_{x}=0$, the new rotation parameters $\ti{a}$ and $\ti{b}$, and rescaled electric charge $\tilde{Q}=\frac{Q}{\kappa^3}$.

Now, let us explore the specific form of the functions $X$ and $Y$ in terms of the new variables to understand the meaning of the constants $\alpha, \ n$ and $\mu_{x}$. First, let us note that, following equation \eqref{eq:rafael_metric_functions}, the functions $\tilde{X}(\tilde{p})$ and $\tilde{Y}(\tilde{r})$ now read

\begin{eqnarray}
    &\ti{X}(\tp)&=\lambda\tp^4-\frac{1-\alpha+\lambda(a^2+b^2)}{\kappa^2}\tp^2+\frac{a^2+b^2+\lambda a^2 b^2+2n}{\kappa^4}-\frac{(a b+ Q\mu_{x})^2}{\kappa^6\tp^2} \ , \\
    &\ti{Y}(\tir)&=\lambda  \tir^4+\frac{1-\alpha+\lambda(a^2+b^2)}{\kappa ^2}\tir^2 +\frac{a^2 b^2\lambda +a^2 +b^2-2 m}{\kappa ^4}+\frac{(a b + Q \mu_{y})^2}{\kappa ^6 \tir^2}
\end{eqnarray}

If we could solve the system of equations
\begin{eqnarray} \notag
    \frac{(a b+ Q\mu_{x})^2}{\kappa^6}&=&\ti{a}^2\ti{b}^2 \ , \\
    \frac{a^2+b^2+\lambda a^2b^2+2n}{\kappa^4}&=&\ti{a}^2+\ti{b}^2+\lambda\ti{a}^2\ti{b}^2 \ , \label{ecus} \\ \notag
    \frac{1-\alpha+\lambda(a^2+b^2)}{\kappa^2}&=&1+\lambda(\ti{a}^2+\ti{b}^2) ,
\end{eqnarray}
then we would conclude that the original metric is actually equivalent, at least locally, to the one with vanishing NUT parameter, $\mu_{x}$ and $\alpha$, and rotation parameters $\ti{a}$ and $\ti{b}$. As we will see, this is actually the case. Notice also that, using the relation $\mu_{y}-\mu_{x}=1$, the function $\ti{Y}(\tir)$ is written
\begin{eqnarray}
    &\ti{Y}(\tir)&=\lambda  \tir^4+\frac{1-\alpha+\lambda(a^2+b^2)}{\kappa ^2}\tir^2 +\frac{a^2 b^2\lambda +a^2 +b^2-2 m}{\kappa ^4}+\frac{(a b + Q \mu_{y})^2}{\kappa ^6 \tir^2}=\\
    &&=\lambda  \tir^4+\left(1+\lambda(\ti{a}^2+\ti{b}^2)\right)\tir^2 +\left(\ti{a}^2 \ti{b}^2\lambda +\ti{a}^2 +\ti{b}^2-\frac{2(m+n)}{\kappa ^4}\right)+\frac{(\ti{a}\ti{b} + \ti{Q})^2}{\tir^2}
\end{eqnarray}
and, then, the mass of the new solution would be $\tilde{m}=\frac{m+n}{\kappa^4}$ and the charge $\ti{Q}=\frac{Q}{\kappa^3}$.

In conclusion, whenever equations \eqref{ecus} can be solved, all metrics in \eqref{eq:metric_rafael} are related to the CCLP solution. Therefore, we must analyze this system carefully. For clarification, we will separate the analysis into the case of vanishing and non-vanishing cosmological constant.

\subsection{The case $\lambda=0$}

If $\lambda=0$, $\alpha\neq{1}$ the polynomial $p^2X(p)$ is degree 2 in $p^2$, and the equations \eqref{ecus} are solved for 
\begin{eqnarray}
    &\kappa=&\sqrt{1-\alpha} \\
    &\tilde{a}^2=&\frac{p_{-}^2}{\kappa^2}=\frac{a^2+b^2+2n-\sqrt{(a^2+b^2+2n)^2-4(1-\alpha)(a b + Q \mu_{x})^2}}{2(1-\alpha)^2}\\
    &\tilde{b}^2=&\frac{p_{+}^2}{\kappa^2}=\frac{a^2+b^2+2n+\sqrt{(a^2+b^2+2n)^2-4(1-\alpha)(a b + Q \mu_{x})^2}}{2(1-\alpha)^2}
\end{eqnarray}
where $p_{-}$ and $p_{+}$ are the roots of $p^2X(p)$, and consequently $X(p)$ is written as:
\begin{equation}
    X(p)=(\alpha-1)\frac{(p^2-p_{-}^2)(p^2-p_{+}^2)}{p^2}.
\end{equation}
The condition $X(p)>0$ together with the existence of real roots for $X(p)$ lead to the condition on $\alpha$
\begin{equation}
    1-\left(\frac{a^2+b^2+2n}{2(a b + Q \mu_{x})}\right)^2<\ \alpha<\ 1 \ .
\end{equation}

Hence, the general metric with $\lambda=0$ is equivalent to the CCLP metric with $\lambda=0$, mass parameter $\frac{m+n}{(1-\alpha)^2}$, electric charge $\frac{Q}{(1-\alpha)^{3/2}}$, and rotation parameters $\ti{a}$ and $\ti{b}$.

\subsection{The case $\lambda\neq{0}$}

Provided we have $\tilde{a}$, $\tilde{b}$ and $\kappa$ exist, then $X(\kappa{\tilde{a}})=X(\kappa{\tilde{b}})=X(\frac{\kappa}{\sqrt{\lambda}})=0$. Consequently $\tilde{a}$, $\ti{b}$ and $\kappa$ can be expressed in terms of the roots of the polynomial $U(p)=p^2X(p)$, which is a degree-6 polynomial in $p$ but is cubic in $p^2$. This feature tells us that every root of $U$ has multiplicity at least two. $X(p)$ can be written as 
\begin{equation} \label{Xfactorized}
    X(p)=\lambda\frac{(p^2-p_0^2)(p^2-p_1^2)(p^2-p_2^2)}{p^2}
\end{equation}
where in general, $p_0^2$, $p_1^2$ and $p_2^2$ are complex numbers. Furthermore, if we set $\tilde{a}$, $\ti{b}$ and $\kappa$ such that
\begin{equation} \label{tildedparams}
    \tilde{a}^2=\frac{p_0^2}{\lambda\ {p_2^2}} \ , \ \tilde{b}^2=\frac{p_1^2}{\lambda\ p_2^2} \ , \ \kappa^2=\lambda \ p_2^2 \ ,
\end{equation}
then the fulfillment of equations \eqref{ecus} is a mere consequence of Vieta's formulas.

Let us notice that, in general, any metric will be locally equivalent to the CCLP metric, although perhaps involving a complex rotation, charge, or mass parameters.

As mentioned above, the polynomial $p^2X(p)$ has only even degree terms in $p$, and consequently it is also a polynomial on variable $t:=p^2$, so we define $V(t)$ such that $V(p^2)=U(p)=p^2X(p)$. This means that:
\begin{equation}
    V(t)=\lambda(t-t_0)(t-t_1)(t-t_2) \ ,
\end{equation}
where $t_0=p_0^2$, $t_1=p_1^2$ and $t_2=p_2^2$.

\subsubsection{Asymptotically Anti-de Sitter spaces: CCLP}

Let us find conditions for the metric to be equivalent to CCLP with real parameters. This would be the case in which the three roots of $V(t)$ are positive.

Rewriting equations \eqref{ecus} in terms of $t_0=p_0^2$, $t_1=p_1^2$ and $t_2=p_2^2$, we get
\begin{eqnarray} \notag
    t_0t_1+t_0t_2+t_1t_2&=&\frac{a^2+b^2+\lambda a^2b^2+2n}{\lambda} \\ \label{ecust1}
    t_0+t_1+t_2&=&\frac{1-\alpha+\lambda(a^2+b^2)}{\lambda}\ , \\
    t_0t_1t_2&=&\frac{(a b+ Q\mu_{x})^2}{\lambda}. \notag
\end{eqnarray}
In order to see exactly when the metrics are equivalent to CCLP we need to solve this equations for positive $t_0,t_1,t_2$.

Notice that this requires $\lambda>0$, $n>-\frac{a^2+b^2+\lambda a^2 b^2}{2}$ and $\alpha<1+\lambda(a^2+b^2)$. These conditions are not sufficient to guarantee the existence of three positive solutions.

Under the change of variables 
\begin{equation}
    \begin{aligned}
    t_0&=&\frac{u_0-\sqrt{2}C u_1}{\sqrt{3}}\ ,\\
    t_1&=&\frac{u_0}{\sqrt{3}}+\frac{C u_1}{\sqrt{6}}+\frac{C u_2}{\sqrt{2}}\ , \\
    t_2&=&\frac{u_0}{\sqrt{3}}+\frac{C u_1}{\sqrt{6}}-\frac{C u_2}{\sqrt{2}}
    \end{aligned}
\end{equation}
where $C=\frac{1-\alpha+\lambda(a^2+b^2)}{\sqrt{3}\lambda}$, the first two equations in \eqref{ecust1} read
\begin{eqnarray} \notag
    u_0^2-\frac{C^2}{2}(u_1^2+u_2^2)&=&\frac{a^2+b^2+\lambda a^2b^2+2n}{\lambda}=K^2 \\ \label{ecust}
    u_0&=&\frac{1-\alpha+\lambda(a^2+b^2)}{\sqrt{3}\lambda}=C .
\end{eqnarray}

Once the second equation is used, the first equation reads
\begin{equation}
    u_1^2+u_2^2=2\left(1-\frac{K^2}{C^2}\right)=:R^2
\end{equation}

The conditions that $t_0$, $t_1$ and $t_2$ are all positive are written in terms of $u_0$, $u_1$ and $u_2$ as follows
\begin{equation}
    u_0=C>0 \ \ , \ \ \sqrt{6} u_1+3 \sqrt{2} u_2+2 \sqrt{3}>0\ \ ,\ \ \sqrt{6}u_1+2 \sqrt{3}>3 \sqrt{2} u_2 \ \ , \ \ \sqrt{2} u_1<1
\end{equation}
which defines an equilateral triangle. If we use the equations and conditions, the third equation in $\eqref{ecust}$ is
\begin{equation}\label{curvahiperb}
   -\frac{\sqrt{2} u_1^3+3 u_1^2-3 \sqrt{2} u_1 u_2^2+3 u_2^2-2}{4 \sqrt{2}}=\frac{F^2}{C^3} \ .
\end{equation}

\begin{figure}[htp]
    \centering
    \includegraphics[width=5cm]{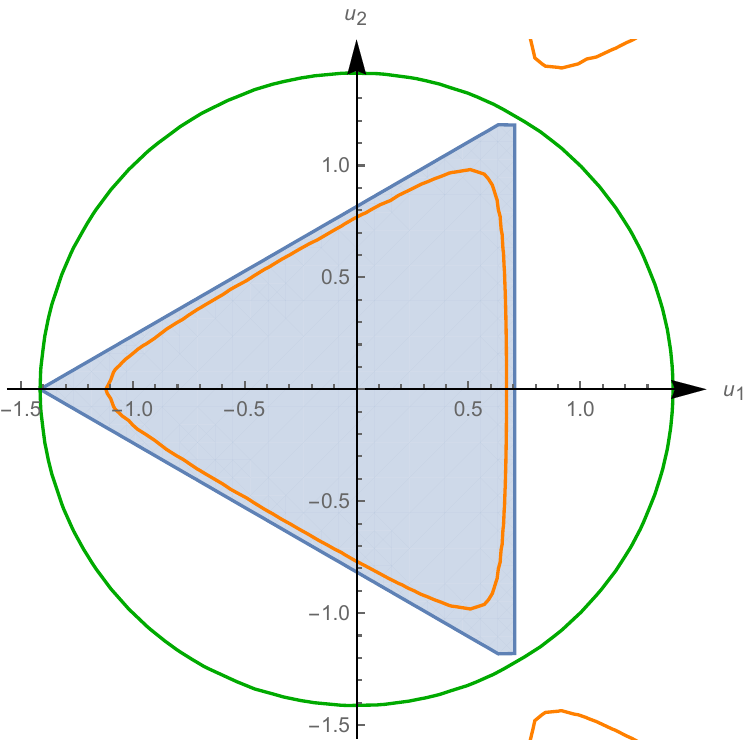}
     \includegraphics[width=5cm]{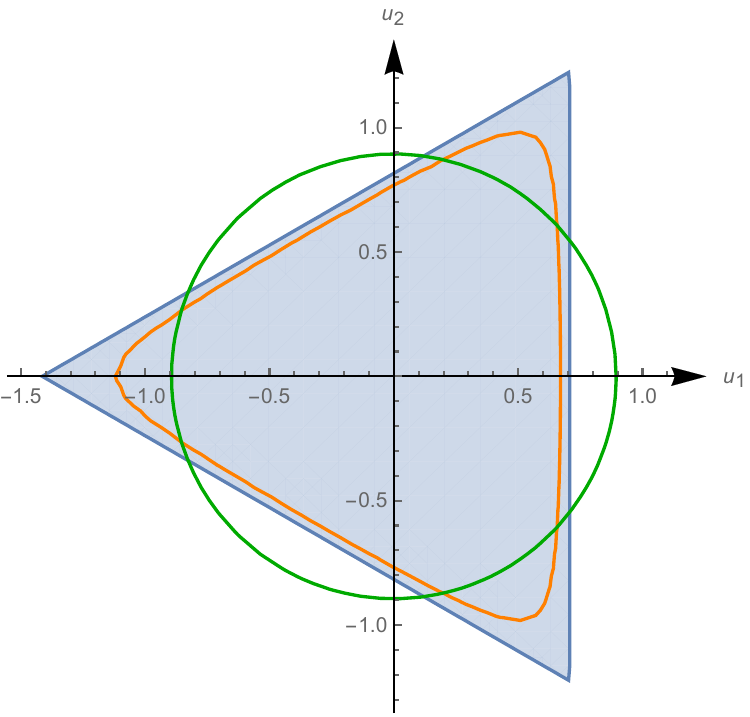}
     \includegraphics[width=5cm]{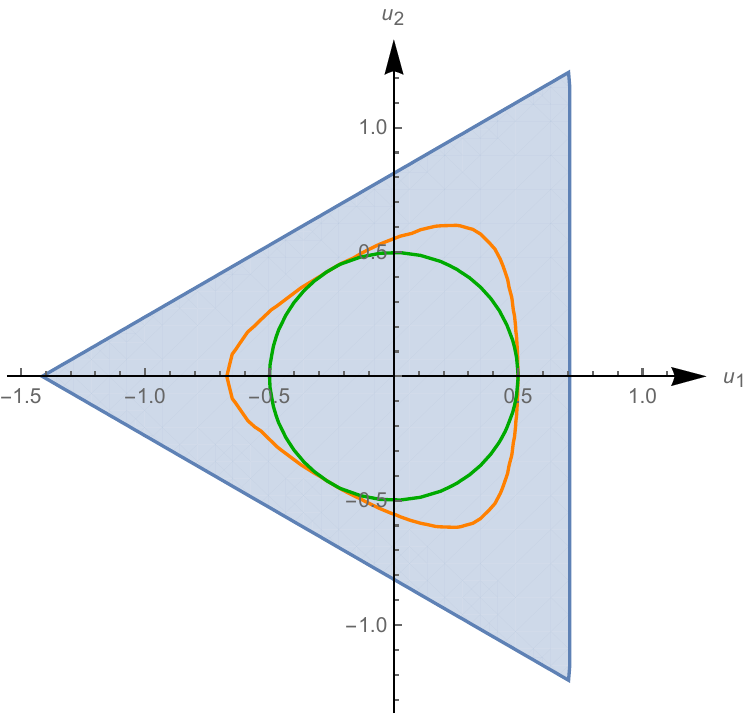}
    \caption{The intersection between $u_1^2+u_2^2=R^2$ (green) and the cubic curve $t_0t_1t_2=F^2$ (orange), in the region $\{t_0>0\wedge{t_1>0}\wedge{t_2>0}\}$(blue). It can be observed that the radius has to be less than $\sqrt{2}$ for there to be a non-empty intersection.}
    \label{fig:intersection}
\end{figure}

Because the curves \eqref{curvahiperb} are symmetric under a rotation of $\frac{2\pi}{3}$, if there is an intersection with the circumference of the radius $R$, then they intersect at least at three points such that at least two of them have different values for the coordinate $u_1$ (see figure \ref{fig:intersection}).

Then, we have to require the discriminant of equation \eqref{curvahiperb}, once $u_2^2=R^2-u_1^2$ is used, to be positive. The resulting cubic equation is
\begin{equation}
   -u_1^3+\frac{3 R^2 u_1}{4}-\frac{3 R^2}{4 \sqrt{2}}+\frac{1}{2 \sqrt{2}}=\left(\frac{3}{2}\right)^{\frac{3}{2}}\frac{F^2}{C^3}
\end{equation}
and its discriminant:
\begin{equation}
    -8 F^4-\frac{8 F^2 \left(3 R^2-2\right)}{3 \sqrt{3}}+\frac{2}{27} \left(R^2-2\right)^2 \left(2 R^2-1\right)>0 \ .
\end{equation}
This leads to the conditions
\begin{eqnarray}
    \frac{3}{4}<\frac{K^2}{C^2}<1 \ \ \text{and} \ \ \frac{3 C K^2-2 C^3-2 \sqrt{\left(C^2-K^2\right)^3}}{3 \sqrt{3}}<F^2<\frac{3 C K^2-2 C^3+2 \sqrt{\left(C^2-K^2\right)^3}}{3 \sqrt{3}}
\end{eqnarray}
or
\begin{eqnarray}
    0<\frac{K^2}{C^2}<\frac{3}{4} \ \ \text{and} \ \ 0<F^2<\frac{3 C K^2-2 C^3+2 \sqrt{\left(C^2-K^2\right)^3}}{3 \sqrt{3}}
\end{eqnarray}

Recalling the definitions of $K$ and $C$, we find
\begin{equation}
    \frac{K^2}{C^2}=\frac{3 \lambda  \left(a^2 b^2 \lambda +a^2+b^2+2 n\right)}{\left(\lambda  \left(a^2+b^2\right)-\alpha +1\right)^2}
\end{equation}

The conditions on $F^2=\frac{(ab+Q\mu_x)^2}{\lambda}$ are rather cumbersome in terms of $\alpha$, $a$, $b$, $\lambda$ and $n$, but the conditions on $\frac{K^2}{C^2}$ ensure the consistency of the bounds for $F^2$ in each case.

\subsubsection{Asymptotically de Sitter spaces}

The asymptotically de Sitter case is obtained when one of the roots is negative and the other two are positive. Let us notice that in such case the function $X(p)$ would also remain positive only for $p\in(p_1,p_2)$ if $p_1^2$ and $p_2^2$ are the two positive roots of $V(t)$, $0<p_1<p_2$.

In this case, we obtain from the third equation in \eqref{ecust} that $\lambda<0$, and the constants $K^2$ and $C$ can have in principle any sign.

For $C>0$ which is equivalent to $\alpha>1+\lambda(a^2+b^2)$ we obtain that
\begin{equation}
    \begin{aligned}
    R>\frac{1}{\sqrt{2}} \iff&& \frac{K^2}{C^2}<\frac{3}{4} \\ 
    -C^3\frac{\sqrt{2}R^3+3R^2-2}{6\sqrt{3}}\leq&&\frac{(a b +Q\mu_x)^2}{\lambda}<0
    \end{aligned}
\end{equation}

\begin{figure}[H]
    \centering
    \includegraphics[width=5cm]{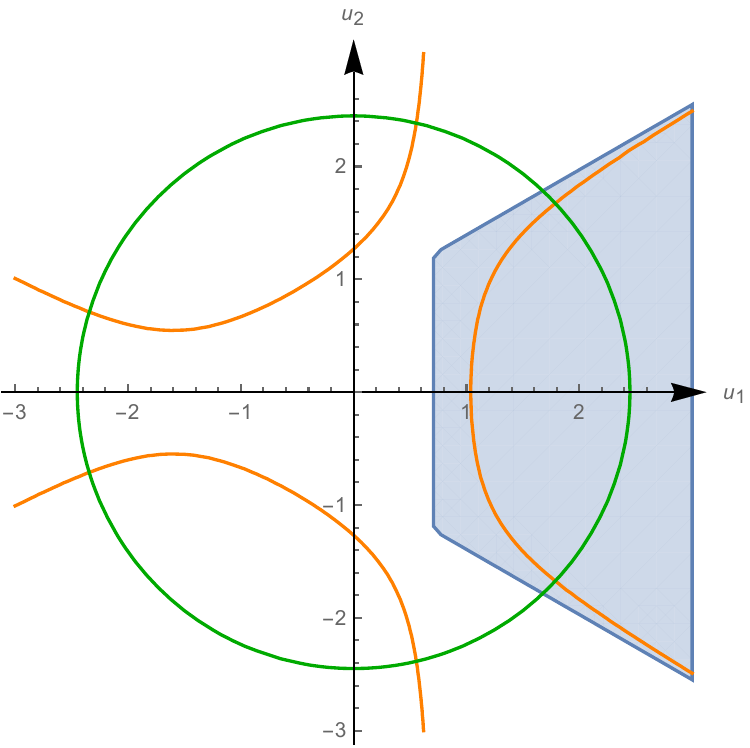}
    \hspace{1cm}
     \includegraphics[width=5cm]{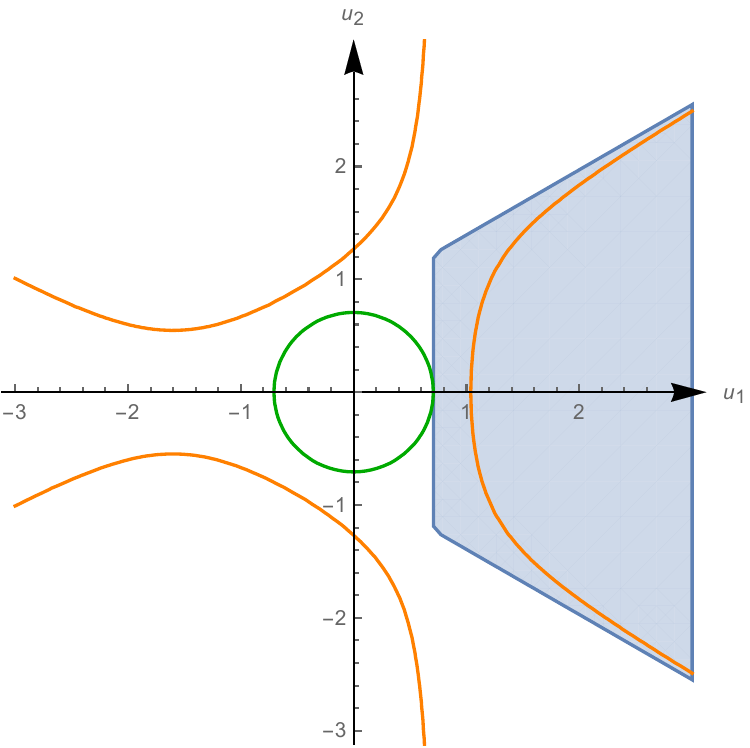}
    \caption{For $\alpha>1+\lambda(a^2+b^2)$, the intersection between $u_1^2+u_2^2=R^2$ (green) and the cubic curve $t_0t_1t_2=F^2<0$ (orange), in the region $\{t_0<0\wedge{t_1>0}\wedge{t_2>0}\}$(blue). In the right figure it can be observed that the radius has to be larger than $\frac{1}{\sqrt{2}}$ for there to be a non-empty intersection.}
\label{fig:intersectionCpositive}
\end{figure}

For $C<0$ which is equivalent to $\alpha<1+\lambda(a^2+b^2)$ we obtain
\begin{equation}
    \begin{aligned}
    R>\sqrt{2} \iff K^2<0 \iff n<\frac{a^2+b^2+\lambda a^2 b^2}{2} \\ 
    C^3\frac{\sqrt{2}R^3-3R^2+2}{6\sqrt{3}}\leq\frac{(a b +Q\mu_x)^2}{\lambda}<0
    \end{aligned}
\end{equation}

Figures 2 and 3 illustrate in a geometrical way the cases in which the system (31,32) has solutions for $\lambda<0$.

\begin{figure}[H]
    \centering
    \includegraphics[width=6cm]{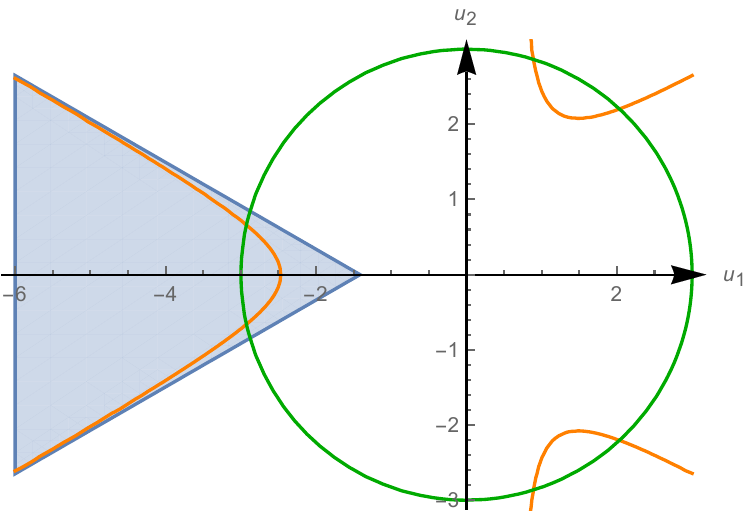}
    \hspace{1cm}
     \includegraphics[width=6cm]{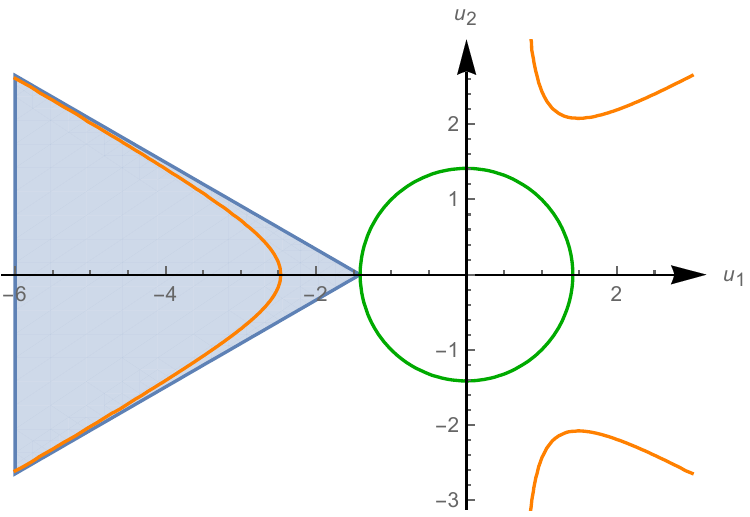}
    \caption{For $\alpha<1+\lambda(a^2+b^2)$, the intersection between $u_1^2+u_2^2=R^2$ (green) and the cubic curve $t_0t_1t_2=F^2<0$ (orange), in the region $\{t_0<0\wedge{t_1>0}\wedge{t_2>0}\}$(blue). In the right figure it can be observed that the radius has to be larger than $\sqrt{2}$ for there to be a non-empty intersection.}
\label{fig:intersectionCnegative}
\end{figure}

\subsection{Komar charges, coordinate transformation and time}\label{sec:komar_charges}

In order to prove the equivalence between the metric with non-vanishing parameters $n$, $\mu_x$ and $\alpha$, and the CCLP metric, we have worked in the coordinate system $(\tilde{t},\tilde{\phi},\tilde{\psi},\tilde{r},\tilde{p})$, in which the components of both metrics are exactly the same. This coordinate system, however, is not asymptotically well behaved globally. To see this, we can reason as follows: the metric that we have is ultimately, CCLP with certain election of parameters, and we know that the asymptotically non-rotating coordinate system is the one in \cite{cclp}. In that coordinate system $n=0=\alpha=\mu_x$ and consequently we do not have any periodicity in the time associated to a NUT parameter. This implies that the correct frame that would give us the correct Komar charges is not that of coordinates $(\tilde{t},\tilde{\phi},\tilde{\psi},\tilde{r},\tilde{p})$, but some other coordinate system in which the metric takes the original CCLP form with parameters $\tilde{a}$, $\tilde{b}$, $\tilde{m}$ and $\tilde{Q}$. In order to find the explicit change of coordinate that brings us there, we notice that had we performed the transformation \eqref{eq:newcoords} and \eqref{eq:newcoordsbis} on the original metric with $\alpha=0=n=\mu_x$ and parameters $\tilde{a}$, $\tilde{b}$, $\tilde{Q}$, we would have arrived to the same metric \eqref{eq:metric_tildedcoords}. As a consequence, the transformation we must perform on \eqref{eq:metric_tildedcoords} to obtain the CCLP metric with new parameters is the inverse of the one previously mentioned.

The full transformation that takes us from the original metric in coordinates $(t,\phi,\psi,r,p)$ to the CCLP metric with parameters $\tilde{a}$, $\tilde{b}$, $\tilde{m}$, $\lambda>0$ and $\tilde{Q}$ in coordinates $(T,\Phi,\Psi,R,P)$ would explicity be given by
\begin{equation}
\begin{aligned}
    t&=\frac{p_2^2(p_2^2-a^2-b^2)+a b (a b + Q \mu_x)}{\sqrt{\lambda}p_2(p_2^2-p_0^2)(p_2^2-p_1^2)}T+\frac{p_0^2(p_0^2-a^2-b^2)+a b (a b + Q \mu_x)}{{\lambda}p_0(p_1^2-p_0^2)(p_2^2-p_0^2)}\Phi\\
    &\qquad-\frac{p_1^2(p_1^2-a^2-b^2)+a b (a b + Q \mu_x)}{{\lambda}p_1(p_1^2-p_0^2)(p_2^2-p_1^2)}\Psi\\[5pt]
    \phi&=\frac{\left(a \lambda  \left(p_2^2-b^2\right) \left(\lambda  p_2^2-1\right)+b \lambda  Q \mu _x\right)}{\lambda ^{3/2} p_2 \left(p_2^2-p_0^2\right) \left(p_2^2-p_1^2\right)}T+\frac{\left(a \left(b^2-p_1^2\right) \left(\lambda  p_1^2-1\right)-b Q \mu _x\right)}{\lambda  p_1 \left(p_0^2-p_1^2\right) \left(p_1^2-p_2^2\right)}\Psi\\
    &\qquad-\frac{\left(a \left(b^2-p_0^2\right) \left(\lambda  p_0^2-1\right)-b Q \mu _x\right)}{\lambda  p_0 \left(p_0^2-p_1^2\right) \left(p_0^2-p_2^2\right)}\Phi\\[5pt]
    \psi&=\frac{\left(b \lambda  \left(p_2^2-a^2\right) \left(\lambda  p_2^2-1\right)+a \lambda  Q \mu _x\right)}{\lambda ^{3/2} p_2 \left(p_2^2-p_0^2\right) \left(p_2^2-p_1^2\right)}T+\frac{\left(-p_1^2 \left(a^2 b \lambda +b\right)+a \left(a b+Q \mu _x\right)+b \lambda  p_1^4\right)}{\lambda  p_1 \left(p_1^2-p_0^2\right) \left(p_1^2-p_2^2\right)}\Psi\\
    &\qquad+\frac{\left(-p_0^2 \left(a^2 b \lambda +b\right)+a \left(a b+Q \mu _x\right)+b \lambda  p_0^4\right)}{\lambda  p_0 \left(p_0^2-p_1^2\right) \left(p_0^2-p_2^2\right)}\Phi\\[5pt]
    r&=\sqrt{\lambda } p_2 R \ \ , \ \ p=\sqrt{\lambda } p_2 P.
\end{aligned}
\end{equation}
For $\lambda=0$ the transformation is different, since $\kappa=\sqrt{1-\alpha}$. This is
\begin{equation}
    \begin{aligned}
        t&=\frac{T}{\sqrt{1-\alpha}}+\frac{\left(p_{-}(p_{-}^2-a^2-b^2)+a b p_{+}\sqrt{1-\alpha}\right)}{(p_{+}^2-p_{-}^2)(1-\alpha)}\Phi-\frac{\left(p_{+}(p_{+}^2-a^2-b^2)+a b p_{-}\sqrt{1-\alpha}\right)}{(p_{+}^2-p_{-}^2)(1-\alpha)}\Psi \ , \\[5pt]
    \phi&=\frac{b p_{+}\sqrt{1-\alpha}-a p_{-}}{(p_{+}^2-p_{-}^2)(1-\alpha)}\Phi-\frac{b p_{-}\sqrt{1-\alpha}-a p_{+}}{(p_{+}^2-p_{-}^2)(1-\alpha)}\Psi ,\\[5pt]
    \psi&=\frac{a p_{+}\sqrt{1-\alpha}-b p_{-}}{(p_{+}^2-p_{-}^2)(1-\alpha)}\Phi-\frac{a p_{-}\sqrt{1-\alpha}-b p_{+}}{(p_{+}^2-p_{-}^2)(1-\alpha)}\Psi \ , \\[5pt]
    r&=\sqrt{1-\alpha}R \ \ , \ \ p=\sqrt{1-\alpha}P \ ,
    \end{aligned}
\end{equation}
and, if we additionally ask for $\mu_x=0$ and $\alpha=0$, this translates into
\begin{equation}
\begin{aligned}
    t&=T+\frac{2 n p_{-}}{p_{+}^2-p_{-}^2}\Phi-\frac{2 n p_{+}}{p_{+}^2-p_{-}^2}\Psi \ , \\[5pt]
    \phi&=\frac{b p_{+}-a p_{-}}{p_{+}^2-p_{-}^2}\Phi-\frac{b p_{-}-a p_{+}}{p_{+}^2-p_{-}^2}\Psi \ \ , \ \ \psi=\frac{a p_{+}-b p_{-}}{p_{+}^2-p_{-}^2}\Phi-\frac{a p_{-}-b p_{+}}{p_{+}^2-p_{-}^2}\Psi \ , \\[5pt]
    r&=R \ \ , \ \ p=P \ .
\end{aligned}
\end{equation}

For $n=0$ then $p_{-}=a$, $p_{+}=b$ and this is the identity transformation, as expected. When $n\neq{0}$ this is the transformation that allows us to remove the NUT parameter in the Kerr-NUT solution.

It is interesting to observe that the original time $t$ was periodic, as the Killing vector generating the null surfaces $p=p_1$, $p=p_2$ necessarily contained a $\partial_{t}$ component\cite{kadsnut}. However, now that we are in the CCLP frame, it is known that the Killing vectors generating the $p=p_1$ and $p=p_2$ null hypersurfaces are precisely $\partial_{\phi}$ and $\partial_{\psi}$ respectively. Consequently, this new time is not periodic. Moreover, the surface gravity of the Killing vectors $\partial_\phi$ and $\partial_\psi$ in these surfaces are euclidean and they are equal to 1, meaning that those variables are periodic with period $2\pi$\cite{kadsnut}.

This new frame is the one that behaves correctly asymptotically, i.e. free of conical singularities and tending to a non rotating reference frame; consequently, it is the appropriate frame to compute the Komar charges. Indeed, the metric is CCLP in its original frame \cite{cclp}, expressed in terms of the variable $p$ instead of $\theta$; therefore, we could compute the Komar charges of this spacetime by simply taking the formulas in \cite{cclp} and replacing $a$, $b$, $Q$ and $m$ by $\tilde{a}=\frac{p_0}{\sqrt{\lambda}p_2}$, $\tilde{b}=\frac{p_1}{\sqrt{\lambda}p_2}$, $\tilde{m}=\frac{m+n}{\kappa^4}=\frac{m+n}{\lambda^2{p_2^4}}$ and $\tilde{Q}=\frac{Q}{\kappa^3}=\frac{Q}{\lambda^{3/2}p_2^6}$, where $p_0$, $p_1$ and $p_2$ are as specified by equations \eqref{Xfactorized} and \eqref{tildedparams}. This can easily be done. However, in order to confirm this result, we can perform an independent calculation of the charges by resorting to an alternative method. In the next section, we will compute the conserved charges of the solution using the near-horizon formalism developed in \cite{Donnay:2016ejv}; see also references therein and thereof.

\section{Near-horizon analysis}\label{sec:near_horizon}

In the previous sections, we have been studying the family of metrics found in \cite{Ferraro} and found that there is a symmetry --which resembles the degeneracy in ($m-n$) of the Kerr-NUT-AdS metrics \cite{kadsnut}-- that allowed to prove that such family is locally equivalent to the CCLP solution \cite{cclp}. Following the near-horizon formalism developed in \cite{Donnay:2016ejv}, adapted to 5 dimensions \cite{ringo, Mei:nearhorizon}, in this section we will the conserved charges associated to the solutions \eqref{eq:metric_rafael} and analyze their thermodynamics.

\subsection{Boundary Conditions}
First, let us review the near-horizon formalism in the case of charged stationary black holes \cite{memoryeffect, sashalucho, meissner}. We consider the following asymptotic conditions in the near horizon region
\begin{equation}
    \begin{aligned}
        g_{vv} & = -2 \kappa \, \rho + \mathcal{O}(\rho^2) \, , \\
        g_{vA} & = \rho \, \theta_A (x^B) + \mathcal{O}(\rho^2) \, , \\
        g_{AB} & = \Omega_{AB}(x^C) + \rho \, \lambda_{AB}(x^C) + \mathcal{O}(\rho^2) \, ,
    \end{aligned}
\end{equation}

\noindent together with the gauge fixing
\begin{equation}
    g_{\rho \rho} = 0, \qquad g_{v \rho} = 1, \qquad g_{A \rho} = 0 \, ,
\end{equation}
where $\rho \in \mathbb R_{\geq0}$ measures the distance from the horizon, the latter being at $\rho = 0$. $v$ is the null coordinate at the horizon, while the three coordinates on the constant-$v$ slides of the horizon are schematically denoted $x^A$, with $A=1,2,3$. As shown in \cite{Donnay:2016ejv}, the Killing vectors that preserve these boundary conditions form an infinite-dimensional algebra. Since here we are dealing with a charged solution, we also need to consider the expansion for the electromagnetic field
\begin{equation}
    \begin{aligned}
        A_v & = A_v^{(0)} + \rho \, A_v^{(1)}(v, x^A) + \mathcal O(\rho^2) \, , \\
        A_B & = A_B^{(0)}(x^A) + \rho \, A_B^{(1)}(v, x^A) + \mathcal O(\rho^2) \, ,
    \end{aligned}
\end{equation}
and the condition $A_\rho = 0$; see \cite{memoryeffect}. These conditions are not expected to change due to the addition of the Chern-Simons term in the 5-dimensional action. The symmetries associated with the asymptotic Killing vector have associated the Noether charges
\begin{equation}
    Q[T, Y^A, U] = - \frac{1}{16 \pi} \int_H d^3x \sqrt{\det \Omega_{AB}} \left( T g_{vv}^{(1)} + Y^A (g_{vA}^{(1)} + 4 A_A^{(0)} A_v^{(1)} ) + 4 U A_v^{(1)}  \right)  + Q_{CS}\, ,\label{Lacargota}
\end{equation}
which follows from the formalism developed in \cite{Barnich:2001jy} applied to the near-horizon region. These charges are the Noether charges associated to the symmetries generated by the asymptotic Killing vectors $T(x^A)\partial_v$, $Y^A(x^B)\partial_A$, with $U(x^A)$ being the function that accompanies the generator of the local gauge transformation; $T(x^A)$ and $Y^A(x^B)$ correspond to the supertranslations and superrotations at the horizon ($H$); see \cite{memoryeffect} for details and conventions. In (\ref{Lacargota}), $Q_{CS}$ stands for the Chern-Simons (CS) contribution to the near horizon charge, which we will explicitly discuss below. $Q_{CS}$ will only contribute to the electric charge evaluated on $H$, but not to the angular momentum \cite{Glennuevo}. Also, it will not contribute to the electric charge $Q_e$ when computed as a flux integral at infinity; see (\ref{Labelow}) below.

\subsection{Near-Horizon Limit}

As discussed in the previous sections, we can get rid of the extra parameters and work with the CCLP metric given by \eqref{eq:metric_rafael} and the functions \eqref{eq:rafael_metric_functions} where we set $n = \alpha = \mu_x = 0$ and $\mu_y = 1$. This metric can be written as
\begin{align}
    ds^2 = \Lambda\of{r, p} \left( - \frac{Y\of{r}}{\Sigma\of{r, p}} dt^2 + \frac{dr^2}{Y\of{r}} + \frac{dp^2}{X\of{p}} \right) + g_{i j} \big( d x^i + \omega^i\of{r, p} \,  dt \big) \big(d x^j + \omega^j\of{r, p} \, dt \big) 
\end{align}
where the indices $i, j = 1, 2$ and the variables corresponds to $x^1 = \phi$ and $x^2 = \psi$, and we defined the functions
\begin{equation}
    \begin{aligned}
        \Lambda(r, p) & = r^2 + p^2 \, , \\[10pt]
        \omega^\phi\of{r,p} & = \frac{\displaystyle \left(a^2+\frac{a b Q}{b^2+r^2}+r^2\right) \left(a \left(1 + r^2 \lambda\right)+\frac{b Q}{b^2+r^2}\right)-\frac{a r^2 Y(r)}{b^2+r^2}}{\displaystyle \left(a^2+\frac{a b Q}{b^2+r^2}+r^2\right)^2-\frac{r^4 Y(r) \left(\frac{a^2 b^2}{p^2}+\frac{a^2 b^2}{r^2}+\frac{X(p)}{\left(1-\lambda  p^2\right)^2}\right)}{\left(b^2+r^2\right)^2}} \, ,\\[10pt]
        \omega^\psi\of{r, p} & = \frac{ \displaystyle \left(\frac{a b Q}{a^2 + r^2} + b^2 + r^2\right) \left(\frac{a Q}{a^2 + r^2} + b \left(1 + r^2 \lambda \right)\right)-\frac{b r^2 Y(r)}{a^2+r^2}}{ \displaystyle \left(\frac{a b Q}{a^2+r^2}+b^2+r^2\right)^2-\frac{r^4 Y(r) \left(\frac{a^2 b^2}{p^2}+\frac{a^2 b^2}{r^2}+\frac{X(p)}{\left(1-\lambda  p^2\right)^2}\right)}{\left(a^2+r^2\right)^2}} \, , \\[10pt]
        \Sigma\of{r, p} & = \frac{\left(\left(a^2+r^2\right) \left(b^2 + r^2\right) + a b Q\right)^2}{r^4} - Y(r) \left(\frac{a^2 \left(1 - b^2 \lambda \right) + b^2 - p^2}{1 - p^2 \lambda} + \frac{a^2 b^2}{r^2}\right) \, .
    \end{aligned}
\end{equation}
and explicitly
\begin{equation}
    \begin{aligned}
        g_{\phi \phi} & = \frac{a \left(p^2-a^2\right)^2 \left(a \left(2 m \left(p^2+r^2\right)-Q^2\right)+2 b Q \left(p^2+r^2\right)\right)}{\left(b^2-a^2\right)^2 \left(1-a^2 \lambda \right)^2 \left(p^2+r^2\right)^2}+\frac{\left(p^2-a^2\right) \left(a^2+r^2\right)}{\left(b^2-a^2\right) \left(1-a^2 \lambda \right)} \, ,\\[10pt]
        g_{\psi \psi} & = \frac{b \left(b^2-p^2\right)^2 \left(2 a Q \left(p^2+r^2\right)+b \left(2 m \left(p^2+r^2\right)-Q^2\right)\right)}{\left(b^2-a^2\right)^2 \left(1-b^2 \lambda \right)^2 \left(p^2+r^2\right)^2}+\frac{\left(b^2-p^2\right) \left(b^2+r^2\right)}{\left(b^2-a^2\right) \left(1-b^2 \lambda \right)} \, ,\\[10pt]
        g_{\phi \psi} & = \frac{\left(p^2-a^2\right) \left(b^2-p^2\right) \left(Q \left(a^2+b^2\right) \left(p^2+r^2\right)+a b \left(2 m \left(p^2+r^2\right)-Q^2\right)\right)}{\left(b^2-a^2\right)^2 \left(1-a^2 \lambda \right) \left(1-b^2 \lambda \right) \left(p^2+r^2\right)^2} \, .
    \end{aligned}
\end{equation}

The metric horizon is located at $r=r_{+}$, the largest root of $Y(r)$. It is easy to check that the functions $\Sigma\of{r,p}$ and $\omega^i\of{r, p}$ take constant values on this surface. Now, we define the following coordinate change
\begin{equation}
    \begin{aligned}
        v & = t + \int_{r_+}^{r} dr' \frac{\sqrt{\Sigma_+}}{Y\of{r'}} \, , \\
        \tilde x^i & = x^i + \int_{r_+}^r dr' \frac{\sqrt{\Sigma_+} \, \omega_+^i}{Y\of{r'}} - \omega_+^i \, v \, ,
    \end{aligned}
\end{equation}
where we defined $\Sigma_+ \equiv \Sigma\of{r_+, p}$ and $\omega_+^i = \omega^i\of{r_+, p}$. In terms of differential forms, they read
\begin{equation}
    \begin{aligned}
        dv & = dt + \frac{\sqrt{\Sigma_+}}{Y\of{r}} dr \, ,\\
        d \tilde x^i & = d x^i + \frac{\sqrt{\Sigma_+}\, \omega_+^i}{Y\of{r}}dr - \omega_+^i \, dv \, .
    \end{aligned}
\end{equation}

In the new coordinates the metric reads
\begin{align}
    ds^2 & = \Lambda\of{r, p} \left( - \frac{Y\of{r}}{\Sigma\of{r, p}} dv^2 + \frac{\Sigma\of{r, p} - \Sigma_+}{Y\of{r} \, \Sigma\of{r, p}} dr^2 + 2 \frac{\sqrt{\Sigma_+}}{\Sigma\of{r, p}}dv dr + \frac{dp^2}{X\of{p}} \right) \nonumber \\
    & \qquad + \, g_{i j} \Big( d \tilde x^i + \big( \omega^i\of{r, p} - \omega_+^i\big) \, dv + \frac{\sqrt{\Sigma_+}}{Y\of{r}} \big( \omega^i\of{r, p} - \omega_+^i\big) dr \Big) \nonumber \\
    & \hspace{32pt} \times \Big( d \tilde x^j + \big( \omega^j\of{r, p} - \omega_+^j\big) \, dv + \frac{\sqrt{\Sigma_+}}{Y\of{r}} \big( \omega^j\of{r, p} - \omega_+^j\big) dr \Big) \, ,
\end{align}
and the induced metric on the horizon ($H$) is
\begin{equation}
    ds^2 |_H = \frac{\Lambda_{+}(p)}{X\of{p}} dp^2 + g_{i j}|_H d \tilde x^i d \tilde x^j \, ,
    \label{eq:horizon_induced_metric}
\end{equation}
where $\Lambda_{+}(p)=\Lambda(r_{+},p)$. At this point we follow a similar procedure as in \cite{booth}: it is easy to check that $\ell = \partial_v$ is a null vector on the horizon and now we look for another null vector $n = n^\mu \partial_\mu$ such that $n \cdot \ell = -1$, then
\begin{equation}
    n = -\frac{1}{2} g_{i j} |_H \, \omega_+^i \omega_+^j \partial_v - \frac{\sqrt{\Sigma_+}}{\Lambda_{+}(p)} \partial_r - \omega_+^i \partial_i \, ,
\end{equation}
and now we are able to compute the first-order components of the metric by considering the variation of the induced horizon metric along the family of geodesics parametrized by $n$. The order zero components of the metric can be read from \eqref{eq:horizon_induced_metric}, and the first order components are
\begin{equation}
    \begin{aligned}
        g_{vv}^{(1)} & = \frac{Y{'}\of{r_+}}{\sqrt{\Sigma_+}}\ , \quad \, g_{vp}^{(1)} = \frac{\partial_p \Lambda}{\Lambda_{+}(p)}|_H \ , \quad \, g_{pp}^{(1)} = -2 r_+ \frac{\Sigma_+^{1/2}}{\Lambda_{+}(p)} \, , \\[5pt] 
        g_{\tilde \phi \tilde \phi}^{(1)} & = -\frac{\Sigma_+^{1/2}}{\Lambda_{+}(p)} \partial_r g_{\tilde \phi \tilde \phi} |_H \, , \quad g_{\tilde \psi \tilde \psi}^{(1)} = -\frac{\Sigma_+^{1/2}}{\Lambda_{+}(p)} \partial_r g_{\tilde \psi \tilde \psi} |_H \, , \quad g_{\tilde \phi \tilde \psi}^{(1)} = -\frac{\Sigma_+^{1/2}}{\Lambda_{+}(p)} \partial_r g_{\tilde \phi \tilde \psi} |_H \, , \\[5pt]
        g_{v \tilde \phi}^{(1)} & = n^r \left( -g_{\phi \phi} \partial_r \omega^\phi - g_{\phi \psi} \partial_r \omega^\psi \right) |_H \, , \quad g_{v \tilde \psi}^{(1)} = n^r \left( -g_{\psi \psi} \partial_r \omega^\psi - g_{\phi \psi} \partial_r \omega^\psi \right) |_H \, , \\[5pt]
        g_{p \tilde \phi}^{(1)} & = \frac{\Sigma_+ \left( g_{\tilde \phi \tilde \phi} (\omega^{\tilde \phi} - \omega_+^{\tilde \phi}) + g_{\tilde \phi \tilde \psi}(\omega^{\tilde \psi} - \omega_+^{\tilde \psi}) \right)}{\Lambda(r,p)^2 Y(r)} |_H \, , \\[5pt]
        g_{p \tilde \psi}^{(1)} & = \frac{\Sigma_+ \left( g_{\tilde \phi \tilde \psi} (\omega^{\tilde \phi} - \omega_+^{\tilde \phi}) + g_{\tilde \psi \tilde \psi}(\omega^{\tilde \psi} - \omega_+^{\tilde \psi}) \right)}{\Lambda(r,p)^2 Y(r)} |_H \, .
    \end{aligned}
\end{equation}
Now, let us turn to the electromagnetic field: in the new coordinates, we have
\begin{equation}
    A = \Big( A_t + A_\phi \, \omega_+^\phi + A_\psi \, \omega_+^\psi \Big) \Big( dv - \frac{\Sigma_+^{1/2}}{Y(r)} dr \Big) + A_{\phi} \, d\tilde \phi + A_{\psi} \, d \tilde \psi \, .
\end{equation}
where the components of $A = A_\mu(r, p) dx^\mu$ can be read from \eqref{eq:ferraro_em_potential}. In order to satisfy the boundary conditions we need the radial component of the electromagnetic potential to vanish, so we perform the gauge transformation
\begin{equation}
    A \to \tilde A = A + d\beta \, , \quad \text{with} \quad \beta(r, p) = \int_{r_+}^{r} dr' \frac{\Sigma_+^{1/2}}{Y(r')}\Big( A_t + A_\phi \, \omega_+^\phi + A_\psi \, \omega_+^\psi \Big) \, .
\end{equation}

In the near horizon limit the zero order components are given by
\begin{equation}
    \begin{aligned}
        A_p^{(0)} & = 0 \, , \\
        A_v^{(0)} & = \frac{Q}{\Sigma_+^{1/2}} \, , \\
        A_{\tilde \phi}^{(0)} & = A_\phi(r_+, p) \, , \\
        A_{\tilde \psi}^{(0)} & = A_\psi(r_+, p) \, , \\
    \end{aligned}
\end{equation}
and the first order are
\begin{equation}
    \begin{aligned}
        A_v^{(1)} & = \frac{2 Q \, r_+}{\Lambda_+^2(p)} \, , \\
        A_{\tilde \phi}^{(1)} & = - \frac{2 a \, Q (p^2 - a^2) \Sigma_+^{1/2} \, r_+}{(b^2 - a^2)(p^2 + r_+^2)^3 \, \Xi_a} \, , \\
        A_{\tilde \psi}^{(1)} & = - \frac{2 b \, Q (b^2 - p^2) \Sigma_+^{1/2} \, r_+}{(b^2 - a^2)(p^2 + r_+^2)^3 \, \Xi_b} \, , \\
        A_{p}^{(1)} & = \frac{Q}{\Sigma_+^{1/2}} \partial_p n^{v}(p) + A_{i}^{(0)} \partial_p n^{i}(p) \, .
    \end{aligned}
\end{equation}

\subsection{Noether Charges}

Let us evaluate the near-horizon charge (\ref{Lacargota}) for the CCLP solution.

\subsubsection{Wald entropy}
The entropy can be read from the zero-mode associated to the asymptotic Killing vector with $T=1$, $Y^A=0$, $U=0$, i.e. the rigid translation along the null direction $v$. This is
\begin{equation}
    Q_{[1,0,0]} = -\frac{1}{16 \pi} \int_H d^3x \sqrt{g_{ij}^{(0)}} g_{vv}^{(1)} = T_{\text{H}} S_{\text{BH}}
\end{equation}
where $T_{\text{H}}$ stands for the Hawking temperature given by 
\begin{equation}
    2 \pi T_{\text{H}} = \kappa = \frac{r_+^4 \left(\lambda  \left(a^2+b^2+2 r_+^2\right)+1\right)-(a b+Q)^2}{r_+ \left(\left(a^2+r_+^2\right) \left(b^2+r_+^2\right)+a b Q\right)} \, .
\end{equation}
This yields the Bekenstein-Hawking entropy
\begin{equation}
    S_{\text{BH}} = \frac{\pi ^2 \left(\left(a^2+r_+^2\right) \left(b^2+r_+^2\right)+a b Q\right)}{2 r_+ \left(1 - a^2 \lambda\right) \left(1 - b^2 \lambda\right)} \, ,
\end{equation}
in agreement with the result reported in \cite{cclp}.

\subsubsection{Electric charge}
The electric charge is given by the Gauss flux integral
\begin{equation}
    Q_e = \frac{1}{4 \pi} \int_{S^3_\infty} \star F\label{Labelow}
\end{equation}
where we are interested in the component $(\star F)_{p \phi \psi}$ since we want to compute it in a 3-dimensional hypersurface where $t$ and $r$ are taken to be constant values. In this calculation, being computed at infinity, $Q_{CS}\sim -\int_{S_\infty^3} F\wedge A$ does not contribute as the integrand decays faster than the volume of $S^3$ at infinity. To compute the charge at infinity we are considering the gauge \eqref{eq:ferraro_em_potential}. When evaluated at the horizon, in contrast, $Q_{CS}$ does contribute giving a piece that leads to match (\ref{Labelow}), cf. \cite{Glennuevo}. More precisely, one finds{\footnote{The factor $\frac{1}{\sqrt{3}}$ is obtained for $k_{CS}=\frac{1}{3\sqrt{3}}$ and $k_{M}=-\frac{1}{2}$ in order to match our action with the bosonic sector of the action of 5D minimal gauged supergravity in \cite{cclp}}}
\begin{equation}
    Q_{CS}= -\frac{1}{\sqrt{3}}\frac{1}{4 \pi} \int_{H} F\wedge{A}.
\end{equation}

On the horizon, we have
\begin{equation}
    (\star F)_{p \phi \psi}|_H = -\frac{2 Q \, p \, r_+^2 \, \Sigma_+^{1/2}}{(b^2 - a^2) \Lambda_+^2(p) \, \Xi_a \Xi_b} \, ,
\end{equation}
which is nothing but minus the product between
\begin{equation}
    \sqrt{\det g_{ij}^{(0)}} = \frac{p \, r_+ \, \Sigma_+^{1/2}}{(b^2 - a^2) \, \Xi_a \Xi_b} \,  \qquad \text{and} \qquad A_v^{(1)} = \frac{2 Q \, r_+}{\Lambda_+^2(p)} \, .
    \label{eq:volume_form_and_Av1}
\end{equation}
Therefore, we find
\begin{equation}
    Q_e =Q[0,0,1]= \frac{1}{4 \pi} \int_H \star F +Q_{CS}= -\frac{1}{4 \pi} \int_H d^3x \sqrt{\det g_{a b}^{(0)}} \, A_v^{(1)} \,+Q_{CS}\, ,
\end{equation}
in agreement with the charge computed at infinity.

\subsubsection{Angular momentum}
In the same way, we can relate the Komar integral associated to the angular momentum, namely
\begin{equation}
    J_A^{i} = - \frac{1}{16 \pi} \int_H d^3x \left(J_K^{i} + J_{EM}^{i} \right) \, ,
\end{equation}
where we have defined
\begin{equation}
    J_K^{i} = \sqrt{\det g_{i j}^{(0)}} \, g_{v i}^{(1)}, \qquad J_{EM}^{i} = \sqrt{\det g_{i j}^{(0)}} \, 4 A_i^{(0)} A_v^{(1)}  \, ,
\end{equation}
being the contributions of the geometry and the electromagnetic field to the angular momentum, respectively. Let us compute the $J^\phi$ explicitly; a similar computation applies to $J^\psi$. For the computation of $J_K^\phi$, we need the volume form induced on the horizon given in \eqref{eq:volume_form_and_Av1} and the first order component
\begin{equation}
    g_{v \phi}^{(1)} = \frac{2 \left(p^2-a^2\right) \left(b Q \left(r_+^2-a^2\right) \left(p^2+r_+^2\right)+a \left(\left(p^2+r_+^2\right) \left(2 m r_+^2-Q^2\right)-Q^2
   r_+^2\right)+r_+^2 \Sigma_+^{1/2} \omega_+^\phi \left(p^2+r_+^2\right)^2\right)}{r_+ \left(a^2-b^2\right) \, \Xi_a \,
   \left(p^2+r_+^2\right)^3}
 \, ,
\end{equation}

It is possible to check that the first contribution $J_K^\phi$ is equal to the Komar density associated to $K^\mu = {\delta^\mu}_\phi$; namely
\begin{equation}
    (\star d K)_{\phi\psi p} |_H = \sqrt{\det g_{i j}^{(0)}} g_{v \phi}^{(1)} \, ,
\end{equation}
so we see that this term leads to reproduce the result of the angular momentum $J^\phi$ obtained in \cite{cclp}. As for the electromagnetic field contribution
\begin{equation}
    J_{EM}^\phi = - \frac{8 \, a \, p \, (p^2 - a^2) \, Q^2 \, \Sigma_+^{1/2} \, r_+^2}{(b^2 - a^2)^2 (p^2 + r_+^2)^3 \, \Xi_a^2 \, \Xi_b} \, ,
\end{equation}
it can be removed by the gauge transformation $A_\phi^{(0)} \to \tilde A_\phi^{(0)} = A_\phi^{(0)} + \Phi_0^\phi$ with
\begin{equation}
    \Phi_0^\phi = -\frac{a \, Q}{2 \Xi_a \, (b^2 + r_+^2)} \, .
\end{equation}
That is,
\begin{equation}
    \int_H d^3x \sqrt{\det g_{ij}^{(0)}} \, 4 \tilde A_\phi^{(0)} A_v^{(1)} = 0 \, .
\end{equation}

In summary, we obtain that the near-horizon analysis reproduces the correct results for the conserved charges.

\section{Conclusions}\label{sec:conclusions}

In this paper, we have studied the family of solutions of 5-dimensional EMCS theory presented in reference \cite{Ferraro}. We have analyzed their local and global properties, along with their conserved charges. This throws light on on the physical relevance of the multiple parameters of the solutions. We have shown that, by means of a local change of coordinates and a complex redefinition of parameters, the geometries constructed in \cite{Ferraro} turn out to be equivalent to the asymptotically (A)dS solutions found in \cite{cclp}. In some cases, the required coordinate change encounters global obstructions, as it can be thought of as a boosts along angular directions that may introduce closed timelike curves. These observations allow us to compute the conserved charges of the general solutions by importing the results of Komar integrals associated to the conserved charges of CCLP metrics \cite{cclp}. In order to confirm these results, we also performed an independent computation of the conserved charges by resorting to the near-horizon method developed in \cite{{Donnay:2016ejv,Mei:nearhorizon}}. The near-horizon computation yields results in agreement. This is the first time that such a calculation of the near-horizon Noether charges is done in EMCS theory in 5 dimensions.

\subsection*{Acknowledgments}

The authors thank Prof. Rafael Ferraro for his valuable comments and guidance throughout this work.
This paper has been partially supported by CONICET (PIP 11220210100111CO) and Universidad de Buenos Aires (UBACyT 20020220100128BA)


\begin{thebibliography}{10}



\bibitem{Pathak:2016vfc}
A.~Pathak, A.~P.~Porfyriadis, A.~Strominger and O.~Varela,
``Logarithmic corrections to black hole entropy from Kerr/CFT,''
JHEP \textbf{04}, 090 (2017)
[arXiv:1612.04833 [hep-th]].


\bibitem{Ferraro}
R.~Ferraro,
``Electrovacuum geometries in five dimensions,''
Phys. Rev. D \textbf{98} (2018) no.12, 124042
[arXiv:1809.01624 [gr-qc]].

\bibitem{cclp}
Z.~W.~Chong, M.~Cvetic, H.~Lu and C.~N.~Pope,
``General non-extremal rotating black holes in minimal five-dimensional gauged supergravity,''
Phys. Rev. Lett. \textbf{95} (2005), 161301
[arXiv:hep-th/0506029 [hep-th]].

\bibitem{knads}
W.~Chen, H.~Lu and C.~N.~Pope,
``General Kerr-NUT-AdS metrics in all dimensions,''
Class. Quant. Grav. \textbf{23} (2006), 5323-5340
doi:10.1088/0264-9381/23/17/013
[arXiv:hep-th/0604125 [hep-th]].

\bibitem{Arcodia:2021fge}
M.~R.~A.~Arcod\'\i{}a and R.~Ferraro,
``Double-extended Kerr\textendash{}Schild form for 5D electrovacuum solutions,''
Gen. Rel. Grav. \textbf{54} (2022) no.10, 130
[arXiv:2109.09497 [gr-qc]].

\bibitem{kadsnut}
W.~Chen, H.~Lu and C.~N.~Pope,
``Kerr-de Sitter black holes with NUT charges,''
Nucl. Phys. B \textbf{762} (2007), 38-54
[arXiv:hep-th/0601002 [hep-th]].



\bibitem{Cvetic:2010jb}
M.~Cvetic, G.~W.~Gibbons, D.~Kubiznak and C.~N.~Pope,
``Black Hole Enthalpy and an Entropy Inequality for the Thermodynamic Volume,''
Phys. Rev. D \textbf{84}, 024037 (2011)
[arXiv:1012.2888 [hep-th]].

\bibitem{Kinney:2005ej}
J.~Kinney, J.~M.~Maldacena, S.~Minwalla and S.~Raju,
``An Index for 4 dimensional super conformal theories,''
Commun. Math. Phys. \textbf{275}, 209-254 (2007)
[arXiv:hep-th/0510251 [hep-th]].

\bibitem{Cabo-Bizet:2018ehj}
A.~Cabo-Bizet, D.~Cassani, D.~Martelli and S.~Murthy,
``Microscopic origin of the Bekenstein-Hawking entropy of supersymmetric AdS$_{5}$ black holes,''
JHEP \textbf{10}, 062 (2019)
[arXiv:1810.11442 [hep-th]].

\bibitem{Dolan:2013ft}
B.~P.~Dolan, D.~Kastor, D.~Kubiznak, R.~B.~Mann and J.~Traschen,
``Thermodynamic Volumes and Isoperimetric Inequalities for de Sitter Black Holes,''
Phys. Rev. D \textbf{87}, no.10, 104017 (2013)
[arXiv:1301.5926 [hep-th]].

\bibitem{Maldacena:2008wh}
J.~Maldacena, D.~Martelli and Y.~Tachikawa,
``Comments on string theory backgrounds with non-relativistic conformal symmetry,''
JHEP \textbf{10}, 072 (2008)
[arXiv:0807.1100 [hep-th]].

\bibitem{Kunduri:2007vf}
H.~K.~Kunduri, J.~Lucietti and H.~S.~Reall,
``Near-horizon symmetries of extremal black holes,'' Class. Quant. Grav. \textbf{24}, 4169-4190 (2007)
[arXiv:0705.4214 [hep-th]].

\bibitem{Cvetic:2010mn}
M.~Cvetic, G.~W.~Gibbons and C.~N.~Pope,
``Universal Area Product Formulae for Rotating and Charged Black Holes in Four and Higher Dimensions,''
Phys. Rev. Lett. \textbf{106}, 121301 (2011)
[arXiv:1011.0008 [hep-th]].

\bibitem{Chow:2008dp}
D.~D.~K.~Chow, M.~Cvetic, H.~Lu and C.~N.~Pope,
``Extremal Black Hole/CFT Correspondence in (Gauged) Supergravities,'' Phys. Rev. D \textbf{79}, 084018 (2009)
[arXiv:0812.2918 [hep-th]].

\bibitem{Kunduri:2006ek}
H.~K.~Kunduri, J.~Lucietti and H.~S.~Reall,
``Supersymmetric multi-charge AdS(5) black holes,''
JHEP \textbf{04}, 036 (2006)
[arXiv:hep-th/0601156 [hep-th]].

\bibitem{Gutowski:2004ez}
J.~B.~Gutowski and H.~S.~Reall,
``Supersymmetric AdS(5) black holes,''
JHEP \textbf{02}, 006 (2004)
[arXiv:hep-th/0401042 [hep-th]].




\bibitem{Gray:2024qys}
F.~Gray, C.~Keeler, D.~Kubiznak and V.~Martin,
``Love symmetry in higher-dimensional rotating black hole spacetimes,''
JHEP \textbf{03} (2025), 036
[arXiv:2409.05964 [gr-qc]].

\bibitem{Deshpande:2024itz}
R.~Deshpande and O.~Lunin,
``Multi-charged geometries with cosmological constant,''
JHEP \textbf{03} (2025), 131
[arXiv:2408.08254 [hep-th]].

\bibitem{Zhao:2024ehs}
P.~Zhao and H.~L\"u,
``Notes on sums over horizons,''
Phys. Rev. D \textbf{110} (2024) no.2, 024028
[arXiv:2407.13079 [gr-qc]].

\bibitem{Larsen:2024fmp}
F.~Larsen and S.~Lee,
``Supersymmetric charge constraints on AdS black holes from free fields,''
JHEP \textbf{09} (2024), 118
[arXiv:2405.17648 [hep-th]].

\bibitem{Ma:2024ynp}
L.~Ma, P.~J.~Hu, Y.~Pang and H.~Lu,
``Effectiveness of Weyl gravity in probing quantum corrections to AdS black holes,''
Phys. Rev. D \textbf{110} (2024) no.2, L021901
[arXiv:2403.12131 [hep-th]].

\bibitem{Iliesiu:2024cnh}
L.~V.~Iliesiu, A.~Levine, H.~W.~Lin, H.~Maxfield and M.~Mezei,
``On the non-perturbative bulk Hilbert space of JT gravity,''
JHEP \textbf{10} (2024), 220
[arXiv:2403.08696 [hep-th]].

\bibitem{Cano:2024tcr}
P.~A.~Cano and M.~David,
``Near-horizon geometries and black hole thermodynamics in higher-derivative AdS$_{5}$ supergravity,''
JHEP \textbf{03} (2024), 036
[arXiv:2402.02215 [hep-th]].

\bibitem{Hassaine:2024mfs}
M.~Hassaine, D.~Kubiznak and A.~Srinivasan,
``Extremal Kerr-Schild form,''
Phys. Rev. D \textbf{111} (2025) no.6, L061502
[arXiv:2411.17805 [gr-qc]].

\bibitem{Deshpande:2024vbn}
R.~Deshpande and O.~Lunin,
``Rotating Einstein-Maxwell black holes in odd dimensions,''
[arXiv:2411.01795 [hep-th]].

\bibitem{Colombo:2025ihp}
E.~Colombo, V.~Dimitrov, D.~Martelli and A.~Zaffaroni,
``Equivariant localization in supergravity in odd dimensions,''
[arXiv:2502.15624 [hep-th]].

\bibitem{David:2025pby}
M.~David and A.~Vekemans,
``Microstates of AdS$_5$ black holes with hypermultiplets,''
[arXiv:2502.10372 [hep-th]].

\bibitem{Barrientos:2025abs}
J.~Barrientos, C.~Charmousis, A.~Cisterna and M.~Hassaine,
``Rotating spacetimes with a free scalar field in four and five dimensions,''
[arXiv:2501.10223 [gr-qc]].

\bibitem{Hale:2024zvu}
T.~Hale, B.~R.~Hull, D.~Kubiz\v{n}\'ak, R.~B.~Mann and J.~Men\v{s}\'\i{}kov\'a,
``New interpretation of the original charged BTZ black hole spacetime,''
[arXiv:2412.04329 [gr-qc]].

\bibitem{Dias:2024edd}
O.~J.~C.~Dias, P.~Mitra and J.~E.~Santos,
``Charged Rotating Hairy Black Holes in AdS$_5 \times S^5$: Unveiling their Secrets,''
[arXiv:2411.18712 [hep-th]].

\bibitem{Horowitz:2024kcx}
G.~T.~Horowitz and J.~E.~Santos,
``Smooth extremal horizons are the exception, not the rule,''
JHEP \textbf{02} (2025), 169
[arXiv:2411.07295 [hep-th]].

\bibitem{Ezroura:2024xba}
N.~Ezroura and F.~Larsen,
``Supergravity spectrum of AdS$_{5}$ black holes,''
JHEP \textbf{12} (2024), 020
[arXiv:2408.11529 [hep-th]].

\bibitem{Budzik:2023vtr}
K.~Budzik, H.~Murali and P.~Vieira,
``Following Black Hole States,''
[arXiv:2306.04693 [hep-th]].

\bibitem{Mao:2023qxq}
Q.~Y.~Mao, L.~Ma and H.~Lu,
``Horizon as a natural boundary,''
Phys. Rev. D \textbf{109} (2024) no.8, 084053
[arXiv:2307.14458 [hep-th]].

\bibitem{David:2023gee}
M.~David, N.~Ezroura and F.~Larsen,
``The attractor flow for AdS$_{5}$ black holes in $ \mathcal{N} $ = 2 gauged supergravity,''
JHEP \textbf{08} (2023), 090
[arXiv:2306.05206 [hep-th]].

\bibitem{Cabo-Bizet:2024gny}
A.~Cabo-Bizet,
``The Schwarzian from gauge theories,''
[arXiv:2404.01540 [hep-th]].






\bibitem{Donnay:2015abr}
L.~Donnay, G.~Giribet, H.~A.~Gonzalez and M.~Pino,
``Supertranslations and Superrotations at the Black Hole Horizon,''
Phys. Rev. Lett. \textbf{116}, no.9, 091101 (2016)
[arXiv:1511.08687 [hep-th]].

\bibitem{Donnay:2016ejv}
L.~Donnay, G.~Giribet, H.~A.~Gonz\'alez and M.~Pino,
``Extended Symmetries at the Black Hole Horizon,''
JHEP \textbf{09} (2016), 100
[arXiv:1607.05703 [hep-th]].

\bibitem{cosmological}
L.~Donnay and G.~Giribet,
``Cosmological horizons, Noether charges and entropy,''
Class. Quant. Grav. \textbf{36} (2019) no.16, 165005
[arXiv:1903.09271 [hep-th]].

\bibitem{cmetric}
A.~Anabal\'on, S.~Brenner, G.~Giribet and L.~Montecchio,
``Closer look at black hole pair creation,''
Phys. Rev. D \textbf{104} (2021) no.2, 024044
[arXiv:2103.05782 [hep-th]].

\bibitem{memoryeffect}
L.~Donnay, G.~Giribet, H.~A.~Gonz\'alez and A.~Puhm,
``Black hole memory effect,''
Phys. Rev. D \textbf{98} (2018) no.12, 124016
[arXiv:1809.07266 [hep-th]].



\bibitem{meissner}
G.~Giribet, J.~La Madrid, L.~Montecchio, E.~R.~de Celis and P.~Schmied,
``Zooming in on the horizon when in its Meissner state,''
JHEP \textbf{05} (2023), 207
[arXiv:2302.14140 [hep-th]].

\bibitem{sashalucho}
S.~Brenner, G.~Giribet and L.~Montecchio,
``Symmetries of magnetized horizons,''
Phys. Rev. D \textbf{103} (2021) no.12, 124006
[arXiv:2103.06983 [hep-th]].


\bibitem{kacmoody}
G.~Giribet and L.~Montecchio,
``Colored black holes and Kac-Moody algebra,''
Phys. Rev. D \textbf{105} (2022) no.6, 064006
[arXiv:2111.08178 [hep-th]].

\bibitem{Mei:nearhorizon}
C.~Shi and J.~Mei,
``Extended Symmetries at Black Hole Horizons in Generic Dimensions,''
Phys. Rev. D \textbf{95} (2017) no.10, 104053
[arXiv:1611.09491 [gr-qc]].

\bibitem{ringo}
G.~Giribet, J.~Laurnagaray and P.~Schmied,
``Probing the near-horizon geometry of black rings,''
Phys. Rev. D \textbf{108} (2023) no.2, 024061
[arXiv:2304.14461 [hep-th]].

\bibitem{Barnich:2001jy}
G.~Barnich and F.~Brandt,
``Covariant theory of asymptotic symmetries, conservation laws and central charges,''
Nucl. Phys. B \textbf{633} (2002), 3-82
[arXiv:hep-th/0111246 [hep-th]].

\bibitem{booth}
I.~Booth,
``Spacetime near isolated and dynamical trapping horizons,''
Phys. Rev. D \textbf{87} (2013) no.2, 024008
[arXiv:1207.6955 [gr-qc]].

\bibitem{Glennuevo}
G.~Barnich and G.~Compere,
``Conserved charges and thermodynamics of the spinning Godel black hole,''
Phys. Rev. Lett. \textbf{95}, 031302 (2005)
[arXiv:hep-th/0501102 [hep-th]].


\end{thebibliography}

\begingroup\raggedright\endgroup

\appendix{}

\end{document}